\documentclass[frozencache,runningheads,a4paper]{llncs}

\usepackage[english]{babel}

\usepackage{parallel}
\usepackage[frozenCacheFileName=cache,fencedCode]{markdown}
\usepackage{hyperref}
\hypersetup{
    colorlinks=true,
    linkcolor=blue,
    filecolor=magenta,
    urlcolor=cyan,
    pdftitle={vdm2uml},
    pdfpagemode=FullScreen,
}
\usepackage{adjustbox}
\usepackage{tcolorbox}
\tcbuselibrary{skins,breakable}
\usepackage{tikz}
\usetikzlibrary{calc}
\definecolor{syntaxgray}{rgb}{0.98,0.96,0.96}
\definecolor{syntaxbordergray}{rgb}{0.55,0.47,0.44}
\newtcolorbox[]{rulebox}[2][]{%
    empty,
    parbox=false,
    noparskip,
    enhanced,
    boxrule=0.7pt,
    no shadow,
    colback=white,
    left=.03in, 
    top=.04in, bottom=.04in,
    before skip=.2in,
    after skip=.2in,
    title = #1,
    colframe = #2
}
\newtcolorbox[]{testsyntaxbox}[3][\linewidth]{%
    empty,
    parbox=false,
    noparskip,
    enhanced,
    boxrule=0pt,
    no shadow,
    colback=syntaxgray,
    left=.03in, 
    top=.03in, bottom=.03in,
    before skip=.2in,
    after skip=.2in,
    #2,
    #3,
    width=#1
}
\newtcolorbox[]{syntaxbox}[1][\linewidth]{%
    empty,
    parbox=false,
    noparskip,
    enhanced,
    boxrule=0pt,
    no shadow,
    colback=syntaxgray,
    left=.03in, 
    top=.03in, bottom=.03in,
    before skip=.2in,
    after skip=.2in,
    borderline north={1.2pt}{0pt}{syntaxbordergray},
    borderline south={1.2pt}{0pt}{syntaxbordergray},
    width=#1
}
\newtcolorbox[]{notopsyntaxbox}[1][\linewidth]{%
    empty,
    parbox=false,
    noparskip,
    enhanced,
    boxrule=0pt,
    no shadow,
    colback=syntaxgray,
    left=.03in, 
    top=.03in, bottom=.03in,
    before skip=.2in,
    after skip=.2in,
    borderline south={1.2pt}{0pt}{syntaxbordergray},
    width=#1
}
\newtcolorbox[]{nobottomsyntaxbox}[1][\linewidth]{%
    empty,
    parbox=false,
    noparskip,
    enhanced,
    boxrule=0pt,
    no shadow,
    colback=syntaxgray,
    left=.03in, 
    top=.03in, bottom=.03in,
    before skip=.2in,
    after skip=.2in,
    borderline north={1.2pt}{0pt}{syntaxbordergray},
    width=#1
}
\newtcolorbox[]{notopnobottomsyntaxbox}[1][\linewidth]{%
    empty,
    parbox=false,
    noparskip,
    enhanced,
    boxrule=0pt,
    no shadow,
    colback=syntaxgray,
    left=.03in, 
    top=.03in, bottom=.03in,
    before skip=.2in,
    after skip=.2in,
    width=#1
}

\begin{document}

\title{Bidirectional UML Visualisation of VDM Models}

\author{Jonas Lund\inst{1} \and  Lucas Bjarke Jensen\inst{1} \and Nick Battle\inst{2}  \and Peter Gorm Larsen\inst{1} \and Hugo Daniel Macedo\inst{1} }
\authorrunning{ }

\institute{
DIGIT, Aarhus University, Department of Engineering, \\
Finlandsgade 22, 8200 Aarhus N, Denmark\\
\email{\{201906201,201907355\}@post.au.dk, \{pgl,hdm\}@ece.au.dk} \and
Independent
\email{nick.battle@acm.org}
}
			
\maketitle
\setcounter{footnote}{0} 
\begin{abstract}

The VDM-PlantUML Plugin enables translations between the text based UML tool PlantUML and VDM++ and has been released as a part of the VDM VSCode extension. This enhances already extensive feature-set of VDM VSCode with support for UML. The link between VDM and UML is thoroughly described with a set of translation rules that serve as the base of the implementation of the translation plugin. This is however still an early rendition of the plugin with limited usability due to the loss of information between translations and a lack of workflow optimisations, which we plan to solve in the future.


\end{abstract}

\section{Introduction}



As computer-based systems grow in complexity, the necessity for applying formal methods increase as well. One such method is the Vienna Development Method (VDM), which has found success in both industrial and academic environments. The object-oriented (OO) dialect of VDM, VDM++, extends it with classes, objects, inheritance, among other object-oriented features allowing for the modeling of complex software systems.  

An abstract visual representation of the architecture of a system can be provided using Unified Modeling Language (UML) \cite{UML251} class diagrams. This visualisation is useful both when communicating the structure of a system and when designing the architecture itself. Because of this, multiple couplings between various VDM and UML tools have been made. However, the open-source IDE for VDM, the Overture tool, previously being based on the Eclipse platform, is now taking steps to becoming a Visual Studio Code (VS Code) extension \cite{Rask&20,Rask&21}, where UML support is yet to be implemented.

This presents the opportunity to develop a coupling with a faster workflow and support for a more thorough representation of VDM components. 
This coupling was proposed in a paper for the Overture Workshop 2022 \cite{Lund&22}
as a VS Code plugin that would enable a bidirectional translation between the open-source text-based diagram tool PlantUML and VDM VSCode.

Since then, the VDM-PlantUML plugin has been developed and released on VDM VSCode. Some changes were made to the proposed architecture, such as omitting the use of class mapping and instead opting for a simpler architecture using a set of visitor patterns.

Progress has been made in the long term goals mentioned in \cite{Lund&22}
, one of these being the implementation of continual updates on the bidirectional mapping between VDM and UML tools. It is now possible to have the VDMJ language server listen for events, such as a change to a model file, and thus perform the translations continually. To circumvent the information loss that happens between translations, in cooperation with the people at PlantUML, progress is being made on an abstract representation that encapsulates both the specifications from the VDM model as well as the PlantUML model. This would allow users to write complex VDM specifications and translate them back and forth between UML and VDM, without loosing any information included in the original VDM specification. 

Another goal mentioned was static analysis of models that would warn users when a VDM construct lacks a compatible UML counterpart, or ensuring that UML models are properly formatted for translation to VDM. 
The need for the former is somewhat alleviated since associative VDM constructs that are too complex to be shown in a UML model are still included in the model, now as class members instead of associations. This is a part of the inherent abstraction in UML.

The need for the latter is somewhat enhanced since PlantUML has a very flexible syntax. It is therefore relatively easy to make choices that do not properly map onto VDM 
and static analysis would therefore be helpful. Progress has been made in this area by defining the subset of PlantUML that describes VDM models formally, called PlantUML-for-VDM, which can be found in this papers appendix in section \ref{sec:langdef}.

\section{Background}

\subsection{VDM++}

VDM++ is an object-oriented dialect in the VDM language family \cite{Durr&92}. 
It builds upon VDM-SL while introducing OO concepts such as classes, making it suitable for visualisation with class diagrams.

Many tools that add to the functionality of VDM have been developed. One of these is the command line tool VDMJ which is built on Java \cite{Battle09}. Some of the features VDMJ provides are type checking, proof obligations, and combinatorial test generation, among others.

For VDM VSCode, the features of VDMJ are accessed through the Language Server Protocol \cite{Rask21}, letting VS Code act as an interface for VDMJ. One way of implementing a new plugin for VDMJ is by creating an analysis plugin. Analysis plugins make use of the VDMJ interpreter to perform analyses on whatever VDM specification is open in VS Code. It also happens to be how other translation features are implemented on VDM VSCode.

\subsection{UML}

UML is an industry standard visual tool for designing and modeling software-based systems \cite{UML251}. It is specified by the Object Management Group (OMG) to provide a semi-formal visual language to give an abstract representation of OO systems. UML is therefore very relevant as an aid in VDM as a tool for designing and communicating the architecture of a system. UML class diagrams in particular help enhance the utility that VDM++ offers, since VDM++ provides an OO feature-set, and class diagrams can visualise how the different objects in a VDM model relate to each other. 
 
The OMG also maintain their own interchange methods with the XML Metadata Interchange (XMI) standard being used to exchange information using the Extensible Markup Language (XML). XMI is often used as a diagram interchange method for UML models but can be used for any meta model that is expressed in the OMG Meta-Object Facility (MOF).



\section{Transformation Rules}
\label{sec:rules}
In the VDM++ book \cite{Fitzgerald&05}, mapping rules between UML class diagrams and VDM++ are presented . These have been updated to describe the new link between PlantUML and VDM++. The rules are generalised enough to describe any link between VDM++ and UML class diagrams. 
After a transformation rule is stated, a table showing how the rule occurs in both PlantUML and VDM++ is presented, along with the visual representation of the PlantUML code. 
In the transformation rule examples, the classes \texttt{A}, \texttt{B} and \texttt{C} are used when the name of the class is relevant, and \texttt{Class} is used otherwise. \texttt{Type} is used to represent any basic type and any uncapitalised string followed by a number, like \texttt{string1,} is the name of a class member, the details of which will be provided in the explanatory text following the rule. For an overview of the subset of PlantUML syntax that describes VDM models, a language manual is provided in the appendix.

\begin{rulebox}[\textbf{Rule 1: Class Declarations}]{black!80}
There is a one-to-one relationship between classes in UML and classes in VDM++.
\end{rulebox}
\noindent

\par
\begin{Parallel}{0.48\textwidth}{0.48\textwidth}
\ParallelRText{
\begin{center}
    \texttt{VDM++}
\end{center}
\begin{syntaxbox}[\linewidth/2 - 0.1in]
\begin{markdown}
```
class A

... 

End A
```
\end{markdown}
\vspace{18pt}
\end{syntaxbox}
}
\ParallelLText{
\begin{center}
    \texttt{PlantUML}
\end{center}
\begin{syntaxbox}[\linewidth/2 - 0.1in]
\begin{markdown}
```
class A{ ... }
```
\end{markdown}
\centering
\includegraphics[width=0.5\textwidth]{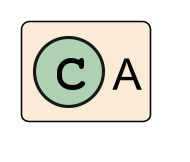}
\end{syntaxbox}
}
\end{Parallel}

\noindent

\noindent
Class members are encapsulated inside class declarations for both the PlantUML and VDM++. However, in PlantUML associations between classes are defined outside class declarations. This means that a VDM++ class member will be defined outside its class in PlantUML, if it is associative. The resulting class diagram shows a single class.

\begin{rulebox}[\textbf{Rule 2: Attributes}]{black!80}
Attributes inside a UML class are represented as instance variables, types or values inside the corresponding VDM++ class.
\end{rulebox}

\par
\begin{Parallel}{0.48\textwidth}{0.48\textwidth}
\ParallelRText{
\begin{center}
    \texttt{VDM++}
\end{center}
\begin{syntaxbox}[\linewidth/2 - 0.1in]
\begin{markdown}
```
instance variables
var1 : Type;
```
\end{markdown}
\vspace{55pt}
\end{syntaxbox}
}
\ParallelLText{
\begin{center}
    \texttt{PlantUML}
\end{center}
\begin{syntaxbox}[\linewidth/2 - 0.1in]
\begin{markdown}
```
var1 : Type

```
\end{markdown}
\centering
\includegraphics[width=0.6\textwidth]{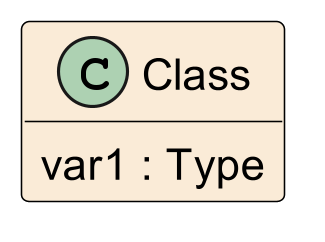}
\end{syntaxbox}
}
\end{Parallel}

\noindent

\noindent
Where \texttt{Type} is the type of the attribute and \texttt{var1} is its identifier. The resulting class diagram shows a class with a single class attribute.

\begin{rulebox}[2.1: Attribute Stereotypes]{black!65}
Instance variables, types and values are differentiated from each other using stereotypes. If no stereotype is used, the attribute is considered an instance variable.
\end{rulebox}
\noindent
\par
\begin{Parallel}{0.48\textwidth}{0.48\textwidth}
\ParallelRText{
\begin{center}
    \texttt{VDM++}
\end{center}
\begin{syntaxbox}[\linewidth/2 - 0.1in]
\begin{markdown}
```
values
val1 : real = value1

types
type1 = nat
```
\end{markdown}
\vspace{41.5pt}
\end{syntaxbox}
}
\ParallelLText{
\begin{center}
    \texttt{PlantUML}
\end{center}
\begin{syntaxbox}[\linewidth/2 - 0.1in]
\begin{markdown}
```
val1 : real «value»
type1 : nat «type»

```
\end{markdown}
\centering
\includegraphics[width=0.92\textwidth]{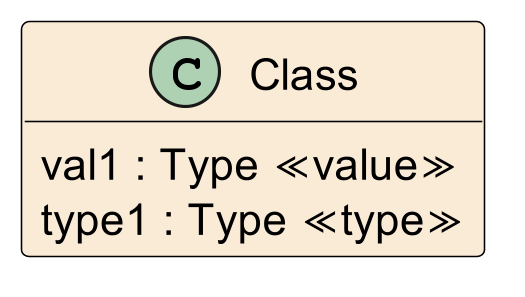}
\end{syntaxbox}
}
\end{Parallel}

\noindent
Where \texttt{val1} and \texttt{type1} are the respective identifiers of the two class members. The stereotype used is the \texttt{attribute stereotype} defined in section \ref{sec:attdef}. The resulting class diagram shows a class containing two attributes with a stereotype denoting a value and a type, respectively.

\begin{rulebox}[\textbf{Rule 3: Functionality}]{black!80}
Operations inside a UML class are represented as functions or operations in the corresponding VDM++ class.
\end{rulebox}

\vspace{25pt}

\begin{Parallel}{0.48\textwidth}{0.48\textwidth}
\ParallelRText{
\begin{center}
    \texttt{VDM++}
\end{center}
\begin{syntaxbox}[\linewidth/2 - 0.1in]
\begin{markdown}
```
operations
op1 : Type ==> Type;
op1() == ( ... );
```
\end{markdown}
\vspace{40.5pt}
\end{syntaxbox}
}
\ParallelLText{
\begin{center}
    \texttt{PlantUML}
\end{center}
\begin{syntaxbox}[\linewidth/2 - 0.1in]
\begin{markdown}
```
op1(Type) : Type
```
\end{markdown}
\centering
\includegraphics[width=0.6\textwidth]{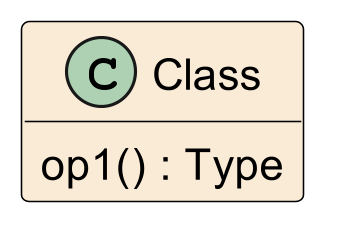}
\end{syntaxbox}
}
\end{Parallel}

\noindent
Where \texttt{op1} is the identifier of the operation. The resulting class diagram shows a class with an operation.

\begin{rulebox}[Rule 3.1: Functional Stereotypes]{black!65}
Operations and functions are differentiated from each other using stereotypes. If no stereotype is used, the attribute is considered an operation.
\end{rulebox}
\par
\begin{Parallel}{0.48\textwidth}{0.48\textwidth}
\ParallelRText{
\begin{center}
    \texttt{VDM++}
\end{center}
\begin{syntaxbox}[\linewidth/2 - 0.1in]
\begin{markdown}
```
functions
func1 : Type ==> Type;
func1() == ( ... );
```
\end{markdown}
\vspace{38pt}
\end{syntaxbox}
}
\ParallelLText{
\begin{center}
    \texttt{PlantUML}
\end{center}
\begin{syntaxbox}[\linewidth/2 - 0.1in]
\begin{markdown}
```
func1() : Type «function»

```
\end{markdown}
\centering
\includegraphics[width=1.05\textwidth]{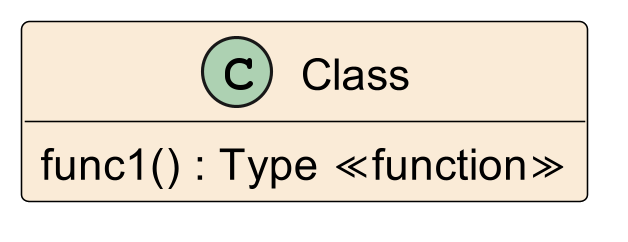}
\end{syntaxbox}
}
\end{Parallel}

\noindent
Where \texttt{func1} is the identifier of the operation. The stereotype used is the \texttt{operation stereotype} defined in section \ref{sec:opdef}. The resulting class diagram shows a class containing an operation with an operation stereotype, denoting a function.

\begin{rulebox}[\textbf{Rule 4: Access modifiers}]{black!80}
All member declarations can have access modifiers at both the UML class diagram; the VDM++ level and the mapping rules for these are one to one, with the exception that undefined and package at the UML class diagram level are considered as private at the VDM++ level. 
\end{rulebox}
\noindent

\par
\begin{Parallel}{0.48\textwidth}{0.48\textwidth}
\ParallelRText{
\begin{center}
    \texttt{VDM++}
\end{center}
\begin{syntaxbox}[\linewidth/2 - 0.1in]
\begin{markdown}
```
private member1
protected member2
public member3
```
\end{markdown}
\vspace{91pt}
\end{syntaxbox}
}
\ParallelLText{
\begin{center}
    \texttt{PlantUML}
\end{center}
\begin{syntaxbox}[\linewidth/2 - 0.1in]
\begin{markdown}
```
- member1
# member2
+ member3
```
\end{markdown}
\centering
\includegraphics[width=0.55\textwidth]{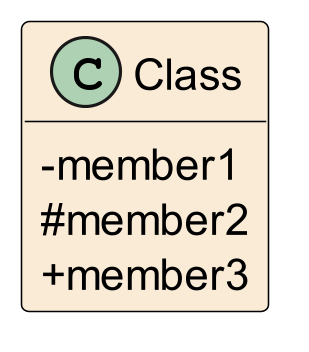}
\end{syntaxbox}
}
\end{Parallel}
\noindent

\noindent
Where \texttt{member1-3} are the identifiers of the three class members. The resulting class diagram shows three class members, each with a different access modifier.

\begin{rulebox}[\textbf{Rule 5: Static definitions}]{black!80}
There is a one-to-one relationship between Static member declarations at the VDM++ level and static member declarations at the UML class diagram level.
\end{rulebox}

\par
\begin{Parallel}{0.48\textwidth}{0.48\textwidth}
\ParallelRText{
\begin{center}
    \texttt{VDM++}
\end{center}
\begin{syntaxbox}[\linewidth/2 - 0.1in]
\begin{markdown}
```
static member1 ...
```
\end{markdown}
\vspace{71.5pt}
\end{syntaxbox}
}
\ParallelLText{
\begin{center}
    \texttt{PlantUML}
\end{center}
\begin{syntaxbox}[\linewidth/2 - 0.1in]
\begin{markdown}
```
{static} member1
```
\end{markdown}
\centering
\includegraphics[width=0.55\textwidth]{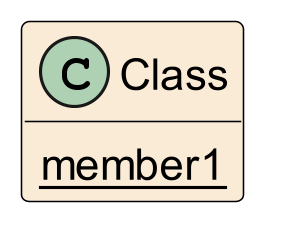}
\end{syntaxbox}
}
\end{Parallel}

\noindent
Where \texttt{member1} is the identifier the class members. The resulting class diagram shows the class member underlined to show that it is static.

\begin{rulebox}[\textbf{Rule 6: Inheritance }]{black!80}
There is a one-to-one relationship between inheritance in UML class diagrams and inheritance in VDM++.
\end{rulebox}

\noindent
\par
\begin{Parallel}{0.48\textwidth}{0.48\textwidth}
\ParallelRText{
\begin{center}
    \texttt{VDM++}
\end{center}
\begin{syntaxbox}[\linewidth/2 - 0.1in]
\begin{markdown}
```
class B is subclass of A

...

end B
```
\end{markdown}
\vspace{0pt}
\end{syntaxbox}
}
\ParallelLText{
\begin{center}
    \texttt{PlantUML}
\end{center}
\begin{syntaxbox}[\linewidth/2 - 0.1in]
\begin{markdown}
```
A <|-- B
```
\end{markdown}
\centering
\vspace{2pt}
\includegraphics[width=0.85\textwidth]{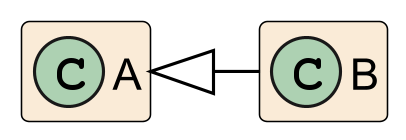}
\end{syntaxbox}
}
\end{Parallel}

\noindent
Here, the resulting class diagram shows how class \texttt{B} inherits from class \texttt{A} using the inheritance arrow head. 

\begin{rulebox}[\textbf{Rule 7: Associations}]{black!80}
Associations between UML classes must be given a role name and a direction. The association is represented in VDM++ as an instance variable in the class at the start of the association, with the role name as its identifier and a value containing a reference to the class at the end of the association.
\end{rulebox}
\noindent

\par
\begin{Parallel}{0.48\textwidth}{0.48\textwidth}
\ParallelRText{
\begin{center}
    \texttt{VDM++}
\end{center}
\begin{syntaxbox}[\linewidth/2 - 0.1in]
\begin{markdown}
```
class A
...
instance variables
assoc1 : B;

end A
```
\end{markdown}
\end{syntaxbox}
}
\ParallelLText{
\begin{center}
    \texttt{PlantUML}
\end{center}
\begin{syntaxbox}[\linewidth/2 - 0.1in]
\begin{markdown}
```
A --> B : assoc1
```
\end{markdown}
\vspace{7pt}
\centering
\includegraphics[width=1.05\textwidth]{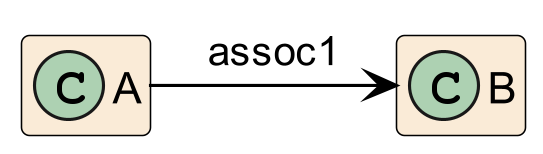}
\end{syntaxbox}
}
\end{Parallel}

\noindent
Where \texttt{assoc1} is the role name of the association and the identifier of the corresponding instance variable. The resulting class diagram shows an association with the direction denoted by the association arrow, with the role name displayed along the link. For a simple association like this, the value of the instance variable is an object reference type.

\begin{rulebox}[\textbf{Rule 8: Association Multiplicity }]{black!80}
The multiplicity and ordering of UML relations determine whether the reference to the class at the end of the association is an object reference type or a compound type with an object reference as its sub-type. 
\end{rulebox}

The different types of multiplicity are describe below in the subrules to rule 8.

\noindent
\label{fig:8.1}
\begin{rulebox}[Rule 8.1: Unordered Collection]{black!65}
An unordered association with a multiplicity of zero-to-many is modelled in VDM++ as a set of object reference types, using the set type. If the multiplicity is one-to-many, the set1 type is used instead.
\end{rulebox}
\noindent

\par
\begin{Parallel}{0.48\textwidth}{0.48\textwidth}
\ParallelRText{
\begin{center}
    \texttt{VDM++}
\end{center}
\begin{syntaxbox}[\linewidth/2 - 0.1in]
\begin{markdown}
```
Class A

instance variables
assoc1 : set of B;
assoc1 : set1 of C;

end A
```
\end{markdown}
\vspace{74pt}
\end{syntaxbox}
}
\ParallelLText{
\begin{center}
    \texttt{PlantUML}
\end{center}
\begin{syntaxbox}[\linewidth/2 - 0.1in]
\begin{markdown}
```
A --> "0..*" B : assoc1
A --> "1..*" C : assoc2
```
\end{markdown}
\centering
\includegraphics[width=1.0\textwidth]{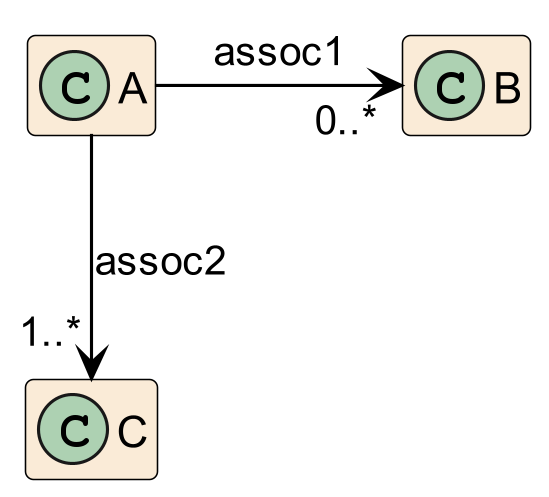}
\end{syntaxbox}
}
\end{Parallel}

\noindent
Where \texttt{assoc1} and \texttt{assoc2} are the role names of the associations between the classes. 
The diagram shows how the association from \texttt{A} to \texttt{B} has a multiplicity of zero-to-many and the association from \texttt{A} to \texttt{C} has a multiplicity of one-to-many. In the VDM++ table, we can see how the \texttt{set} and \texttt{set1} types are used to represent the multiplicity and ordering of the association.

\begin{rulebox}[Rule 8.2: Ordered Collection]{black!65}
An ordered association with a multiplicity of zero-to-many is modelled in VDM++ as a sequence of object reference types, using the \texttt{seq} type. If the multiplicity is one-to-many, the \texttt{seq1} type is used instead.
\end{rulebox}

\par
\begin{Parallel}{0.48\textwidth}{0.48\textwidth}
\ParallelRText{
\begin{center}
    \texttt{VDM++}
\end{center}
\begin{syntaxbox}[\linewidth/2 - 0.1in]
\begin{markdown}
```
Class A

instance variables
assoc1 : seq of B;
assoc2 : seq1 of C;

end A

```
\end{markdown}
\vspace{68.5pt}
\end{syntaxbox}
}
\ParallelLText{
\begin{center}
    \texttt{PlantUML}
\end{center}
\begin{syntaxbox}[\linewidth/2 - 0.1in]
\begin{markdown}
```
A --> "(0..*)" B : assoc1
A --> "(1..*)" C : assoc2

```
\end{markdown}
\centering
\includegraphics[width=1\textwidth]{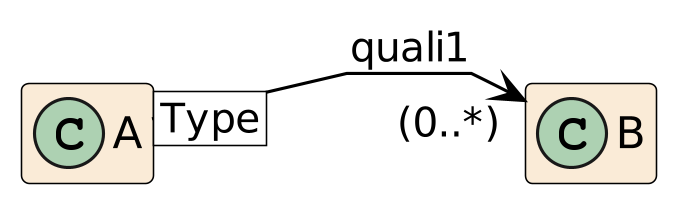}
\end{syntaxbox}
}
\end{Parallel}

\noindent
As in rule 8.1, \texttt{assoc1-2} are the role names of associations between classes, but here the \texttt{seq} and \texttt{seq1} type is used to represent the zero-to-many and one-to-many multiplicity instead, since the collection is now ordered. In PlantUML-for-VDM, an ordered collection is specified by parenthesising the multiplicity. This notation deviates from the UML standard of writing \texttt{\{ordered\}} after the multiplicity. We choose to use this notation for the brevity of it, as the label is more likely to overlap with other elements if it contains a lot of text. 

\begin{rulebox}[Rule 8.3: Optional Association]{black!65}
Any association with a multiplicity of "0..1" is modelled in VDM++ as an optional type with an object reference type as the sub-type.
\end{rulebox}

\noindent
\par
\begin{Parallel}{0.48\textwidth}{0.48\textwidth}
\ParallelRText{
\begin{center}
    \texttt{VDM++}
\end{center}
\begin{syntaxbox}[\linewidth/2 - 0.1in]
\begin{markdown}
```
Class A

instance variables
assoc1 : [B];

end A
```
\end{markdown}
\end{syntaxbox}
}
\ParallelLText{
\begin{center}
    \texttt{PlantUML}
\end{center}
\begin{syntaxbox}[\linewidth/2 - 0.1in]
\begin{markdown}
```
A --> "0..1" B : assoc1
```
\end{markdown}
\vspace{8.7pt}
\centering
\includegraphics[width=1\textwidth]{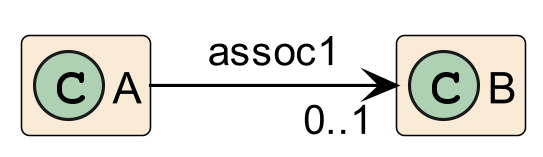}
\end{syntaxbox}
}
\end{Parallel}

\noindent
Since the optional association is at most a single element and therefore not a collection, ordering is not considered.

\begin{rulebox}[\textbf{Rule 9: Qualified Associations }]{black!80}
Qualified associations use a type (or a class name) as the qualifier and are modelled as an instance variable containing an association from the qualifier type to an end sub-type using the map type. 
\end{rulebox}
\noindent

\par
\begin{Parallel}{0.48\textwidth}{0.48\textwidth}
\ParallelRText{
\begin{center}
    \texttt{VDM++}
\end{center}
\begin{syntaxbox}[\linewidth/2 - 0.1in]
\begin{markdown}
```
class A

instance variables
quali1 : map Type to B;

end A
```
\end{markdown}
\vspace{11.5pt}
\end{syntaxbox}
}
\ParallelLText{
\begin{center}
    \texttt{PlantUML}
\end{center}
\begin{syntaxbox}[\linewidth/2 - 0.1in]
\begin{markdown}
```
A [Type] --> B : quali1
```
\end{markdown}
\centering
\includegraphics[width=1.05\textwidth]{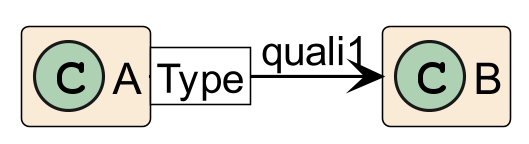}
\vspace{14pt}
\end{syntaxbox}
}
\end{Parallel}

\noindent
Where \texttt{quali1} is the role name of the association and \texttt{Type} is the qualifier for the association. The resulting class diagram shows a qualified association indicated by showing the qualifier in a white box outside the class at the start of the association.

\begin{rulebox}[Rule 9.1: Unique Association Ends]{black!65}
If the mapping from the qualifier to the end sub-type is unique, the association is modelled as an injective mapping instead, using the 'inmap' type.
\end{rulebox}

\noindent

\par
\begin{Parallel}{0.48\textwidth}{0.48\textwidth}
\ParallelRText{
\begin{center}
    \texttt{VDM++}
\end{center}
\begin{syntaxbox}[\linewidth/2 - 0.1in]
\begin{markdown}
```
class A

instance variables
quali1 : inmap Type to B;

end A
```
\end{markdown}
\end{syntaxbox}
}
\ParallelLText{
\begin{center}
    \texttt{PlantUML}
\end{center}
\begin{syntaxbox}[\linewidth/2 - 0.1in]
\begin{markdown}
```
A [(Type)] -> B : quali1
```
\end{markdown}
\centering
\vspace{7.8pt}
\includegraphics[width=1.05\textwidth]{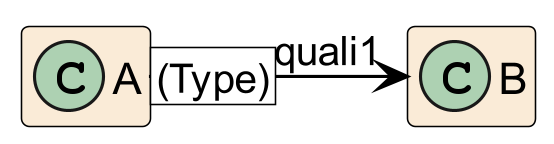}
\end{syntaxbox}
}
\end{Parallel}

\noindent
A qualified association with unique association ends is differentiated from a regular qualified association in PlantUML by parenthesising the qualifier. This deviates from the UML standard of using \texttt{\{unique\}} after the qualifier. We choose to use the parentheses to keep the qualifier label smaller as to not have it overlap with other elements of the diagram.

\noindent
\begin{rulebox}[Rule 9.2: Multiplicity in qualified Associations]{black!65}
For qualified associations, the reference to the class at the end of the association is still subject to rule 8, however the reference in the resulting instance variable in VDM++ is located in the second sub-type of the map type. 
\end{rulebox}

\par
\begin{Parallel}{0.48\textwidth}{0.48\textwidth}
\ParallelRText{
\begin{center}
    \texttt{VDM++}
\end{center}
\begin{syntaxbox}[\linewidth/2 - 0.1in]
\begin{markdown}
```
class A

instance variables
quali1 : inmap Type to 
                    seq of B;

end A
```
\end{markdown}
\end{syntaxbox}
}
\ParallelLText{
\begin{center}
    \texttt{PlantUML}
\end{center}
\begin{syntaxbox}[\linewidth/2 - 0.1in]
\begin{markdown}
```
A [Type] -> "(0..*)" B 
                    : quali1
```
\end{markdown}
\centering
\includegraphics[width=1.08\textwidth]{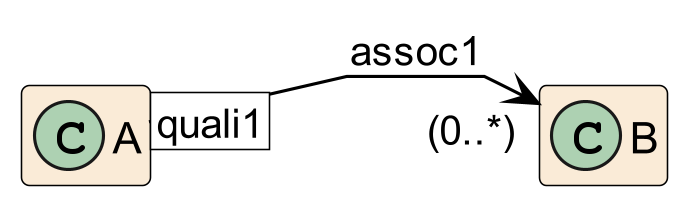}
\end{syntaxbox}
}
\end{Parallel}

\noindent

\noindent

\section{Implementation}


The plugin that translates to  and from PlantUML is implemented as an analysis plugin \cite{analysis_plugin} which operates using the language server of VDMJ \cite{Rask21}. Implementing the translation as an analysis plugin allows for reusing the \texttt{TranslateFeature} which is how other translations such as VDM-to-Word or VDM-to-LaTeX are defined on the client side. It is then as simple as adding the plugin jar to the client's resources and adding VDM-to-UML and UML-to-VDM to the list of possible translations of the \texttt{TranslateFeature}.

Another possible implementation using the VDMJ framework would have been as a command plugin. The translation would then only be callable from a VDM debug console which is more hectic to use than simply pressing a button.

\begin{figure}[h!]
    \begin{adjustbox}{center}
    \includegraphics[width=0.85\textwidth]{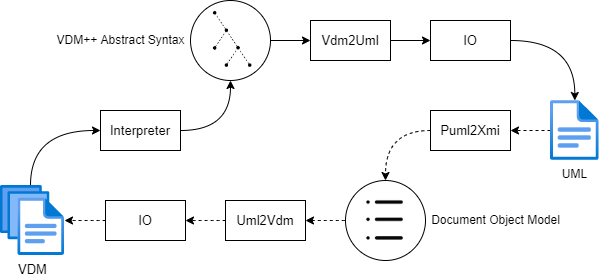}
    \end{adjustbox}
    \caption{Architecture of the translations. The dashed line shows the direction from UML to VDM and the solid line is VDM to UML.}
    \label{fig:archi}
\end{figure}

In Fig. \ref{fig:archi}, the architecture of the plugin is shown. The components implemented are the Vdm2Uml and Uml2Vdm components.

\newpage
\subsection{VDM-to-UML}

The translation to PlantUML can process either an entire workspace folder or a class from a single file. Since the TypeChecker (TC) interpreter is running for the currently active workspace, a single file can only be handled by checking whether the selected URI is a path or not, then assume that the \texttt{UMLGenerator} should be applied to a class with the same name as the file, meaning that single files can only be handled under the condition that a class it contains must have the exact name as the file containing it.

The translation from VDM to UML is implemented using a couple of different TC visitors. One is for traversing classes and their definitions, these being the operations, variables, etc. of the class. What cases are needed is chosen based on what concepts are desired to be shown in the UML output.

The types of the definitions, such as the type of a variable or the return type of an operation, do exist in the nodes of the definition visitor. However, when using these, the types are printed in a way that is undesirable for UML, being either too long or formatted incorrectly for the case. For types another visitor is then used called the \texttt{UMLTypeVisitor} which extends the functionality of the \texttt{TCLeafTypeVisitor} class. This visitor pattern can traverse the nodes of a VDM type and process them to our liking. 

Following the PlantUML formatting details described in section \ref{sec:rules}, classes, the attributes within the classes, and associations between classes are written to string buffers and used as output for the final \texttt{.puml} file. Since the associations are written separately from the classes and their attributes, two string buffers are used to distinguish these definitions.

\subsection{UML-to-VDM}

The translation from UML to VDM uses XMI as an intermediate format between PlantUML and the VDMJ analysis plugin. This is however a tighter link that previous VDM-UML couplings and is not meant to receive XMI from any other source than PlantUML. \\
\noindent

\noindent
The contents of the XML file are parsed using a DOM Parser, 
and the three main elements, the \texttt{UML:Class} element, the \texttt{UML:Generalisation} element, and the \texttt{UML:Association} element are extracted from the XML file. The \texttt{UML:Class} element has two kinds of XML attributes, namely \texttt{UML:Attribute} and \texttt{UML:Operation}. These map directly on to the \texttt{VDM:Attribute}, and \texttt{VDM:Operation} components, with stereotypes being used to differentiate between the sub-components. The \texttt{UML:Generalisation} element has two XML attributes that contain a unique class-ID's of the child and parent class, respectively. The XMI information is mapped onto the \texttt{VDM:Class}, \texttt{VDM:Attribute}, and \texttt{VDM:Operation} components following the mapping rules described in section \ref{sec:rules}. An overview of the mapping of XMI to VDM constructs, can be seen in the figure below.

\begin{figure}[h!]
    \begin{adjustbox}{center}
    \includegraphics[width=0.75\textwidth]{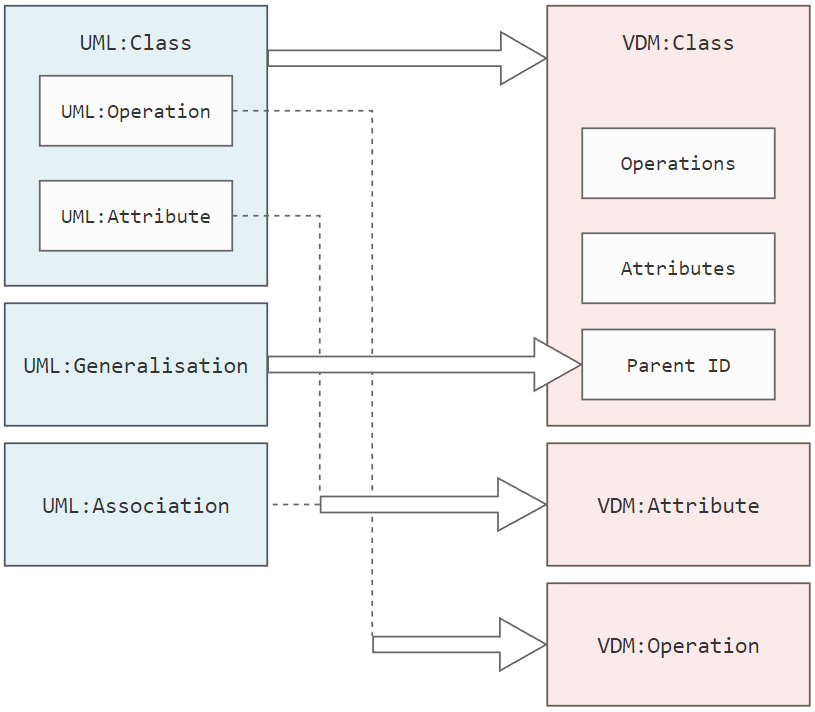}
    \end{adjustbox}
    \caption{Overview of UML2VDM mapping from XMI to VDM constructs}
    \label{fig:uml2vdm}
\end{figure}



\newpage
\section{Conclusion}
\label{sec:conclusion}
As of 12 October 2022, the \verb|vdm-plantuml-plugin| has been offered to users in the newest VDM VSCode release and has already found a use in VDM-specific university courses. The plugin now enables VDM VSCode users to swiftly jump between VDM++ models and visually appealing UML models without dealing with difficult-to-manage file formats and the fuller support of VDM specific components in UML, such as types and functions, enriches the overall VDM experience.

The coupling to PlantUML also increases the interoperability of VDM since PlantUML, as a popular diagram tool, is itself integrated into a large number of tools \cite{running}.
Furthermore, the door have been opened to further collaboration with the PlantUML team, which may lead to new ways of using UML in a VDM context \ref{sec:future}.  

The nine transformation rules between UML and VDM presented in this paper abstracts the PlantUML link to be applicable to any number of UML tools that support the sufficient VDM components. The PlantUML-for-VDM language manual serves as a guide to writing PlantUML that correctly maps to VDM, but is also as a starting point for constructing a syntax checker for creating VDM models in PlantUML.

One significant drawback of using PlantUML is that the graphical visualisation software GraphViz \cite{graphviz}, which PlantUML uses for class diagrams, does not support manually manipulating the positions of the placed classes or the connections between them by clicking and dragging. There are limited ways of changing the layout of a PlantUML diagram, however it would be more ideal to have something more granular to make the diagrams more custom and allow for better workflow and user experience.

\subsection{Future Work}
\label{sec:future}
\subsubsection{VDM-RT/VDM-SL Support}

The plugin was developed and tested with the VDM++ as the main dialect. This is because the object-oriented nature of VDM++ lends itself well to visualisation of the classes a VDM++ project consists of. While VDM-RT transformations are possible, VDM-RT specific constructs such as threads are yet to have their own representation in the plugin. Furthermore, there are plans to allow support for VDM-SL transformations, as a VDM-SL specification can be represented as a singular class. 

\subsubsection{Implicit Stereotypes}

Strides have been made to hide the use of stereotypes from the user, and instead use a definition block structure, mirroring what is done in VDM to group and denote definition statements. 
A successful implementation of the definition block structure in PlantUML-for-VDM would still utilise stereotypes in the XML exchange between PlantUML and VDM VSCode, so no change in the transformation rules will be necessary. 

\subsubsection{Dynamic bidirectionally}

In its current the state, the plugin is suited for presenting the structure of large VDM projects or giving a project a head start by first doing a visual UML model and converting it VDM afterwards. These use cases are mostly one time uses per project and does not bring UML translations to their full potential, workflow-wise. The next step for the link between VDM and UML is the ability to have dynamic bidirectional transformations, meaning that you can work on either the VDM or the UML model and see the effects of those changes immediately take place in the other project. This requires significant changes to the architecture of the link. Progress however is already being made on VDMJ, where a functionality for reacting to certain events using \verb|EventHub| \cite{event_hub} is coming along.

This functionality will also require a new representation of VDM that can help decide what should and should not be translated to UML and retain the parts that are not translated. Doing so will allow the user to write more complicated models while keeping the simplicity of the UML representation.

\subsubsection{Static analysis of VDM-UML}

Static analysis in PlantUML-for-VDM would increase the ease of use when creating a PlantUML model that is to be compliant with the PlantUML-for-VDM language manual, and therefore fit for translation to VDM.

\subsubsection{Coupling with SysML v2}

SysML is Systems Modeling Language. It is a graphical modeling language similar to UML but tailored for systems modeling. SysML v2 is in on-going development and looks promising for coupling with VDM. Thus, it would make sense to look into a coupling that would allow the plugin to also translate a VDM to model to a SysML one.

\section*{Acknowledgements}

We acknowledge the Poul Due Jensen Foundation that funded our basic research for engineering of digital twins.
 
\bibliographystyle{splncs03}
\bibliography{refs.bib, au.bib}

\newcommand{\noop}[1]{}
\begin{thebibliography}{10}
\providecommand{\url}[1]{\texttt{#1}}
\providecommand{\urlprefix}{URL }

\bibitem{running}
Running, \url{https://plantuml.com/running}

\bibitem{graphviz}
Graphviz documentation (September 2022),
  \url{https://graphviz.org/documentation/}

\bibitem{Battle09}
Battle, N.: {VDMJ User Guide}. Tech. rep., Fujitsu Services Ltd., UK (2009)

\bibitem{event_hub}
Battle, N.: Eventhub class on the vdmj github (December 2022),
  \url{https://github.com/nickbattle/vdmj/blob/master/lsp/src/main/java/workspace/EventHub.java}

\bibitem{analysis_plugin}
Battle, N.: Lsp plugin writers guide (January 2023),
  \url{https://github.com/nickbattle/vdmj/wiki/LSP-Plugin-Writers-Guide}

\bibitem{Durr&92}
D{\"u}rr, E., Katwijk, J.: {VDM++, A Formal Specification Language for Object
  Oriented Designs}. In: {COMP EURO 92}. pp. 214--219. IEEE (May 1992)

\bibitem{Fitzgerald&05}
Fitzgerald, J., Larsen, P.G., Mukherjee, P., Plat, N., Verhoef, M.: {Validated
  Designs for Object--oriented Systems}. Springer, New York (2005),
  \url{http://overturetool.org/publications/books/vdoos/}

\bibitem{Lund&22}
Lund, J., Jensen, L.B., Macedo, H.D., Larsen, P.G.: {Towards UML and VDM
  Support in the VS Code Environment}. In: Macedo, H.D., Pierce, K. (eds.)
  Proceedings of the 20th International Overture Workshop. pp. 51--66. Overture
  (7 2022)

\bibitem{UML251}
OMG: About the unified modeling language specification version 2.5.1 (december
  2017), \url{https://www.omg.org/spec/UML/2.5.1}

\bibitem{Rask21}
Rask, J., Madsen, F., Battle, N., Macedo, H., Larsen, P.: The specification
  language server protocol: A proposal for standardised lsp extensions (05
  2021)

\bibitem{Rask&21}
Rask, J.K., Madsen, F.P., Battle, N., Macedo, H.D., Larsen, P.G.: {The
  Specification Language Server Protocol: A Proposal for Standardised LSP
  Extensions}. In: Proen\c{c}a, J., Paskevich, A. (eds.) {\rm Proceedings of
  the 6th Workshop on} Formal Integrated Development Environment, {\rm Held
  online, 24-25th May 2021}. Electronic Proceedings in Theoretical Computer
  Science, vol. 338, pp. 3--18. Open Publishing Association

\bibitem{Rask&20}
Rask, J.K., Madsen, F.P., Battle, N., Macedo, H.D., Larsen, P.G.: {Visual
  Studio Code VDM Support}. In: Fitzgerald, J.S., Oda, T. (eds.) Proceedings of
  the 18th International Overture Workshop. pp. 35--49. Overture (December
  2020), \url{https://arxiv.org/abs/2101.07261}

\end{thebibliography}

\section{Appendix}

\subsection{PlantUML-for-VDM Language Manual}
\label{sec:langdef}

This appendix contains formal definitions for the subset of PlantUML that describes VDM++ models, called PlantUML-for-VDM. At the moment, only the VDM++ is covered, but this will be expanded upon, as the other dialects are supported in the VDM-PlantUML plugin. The purpose of the manual is to serve as a guide for writing PlantUML that can be successfully converted into a VDM model and to establish the proper syntax, for future developments.   

\subsection{Class Declarations}

\begin{syntaxbox}
\begin{markdown}
```
class = ‘class’, identifier, [ class body ], 
                                    [ inheritance clause ] ;

class body = ‘{’, [ definition block ], ‘}’ ;
			
definition block = attribute definitions
                 | functional definitions ;	
                 
inheritance clause = ‘is subclass of’, identifier ;
```
\end{markdown}
\end{syntaxbox}
\subsection{Attribute Definitions}
\label{sec:attdef}

\begin{syntaxbox}
\begin{markdown}
```
attribute definition = instance variable & types
		             | values ;

instance variable & types = [ access member definition ], 
                identifier, ‘:’, type, [ ‘<<type>>’ ] ;

values = [ visibility ], identifier, ‘:’, type, ‘<<value>>’ ;

```
\end{markdown}
\end{syntaxbox}
\noindent
Values are defined separately from instance variable and types, since it does not use the static keyword.

\subsection{Functional Definitions}
\label{sec:opdef}

\begin{syntaxbox}
\begin{markdown}
```
functional definition = operation definition 
                      | function definition ;

operation definition = [ access member definition ], 
            identifier, ‘(’, [ type ], ‘)’,  ‘:’, type' ;

function definition = [ visibility ], identifier, 
        ‘(’, [ type ], ‘)’,  ‘:’, type', ‘<<function>>’ ;

``` 
\end{markdown}
\end{syntaxbox}
\noindent
Where \texttt{type} is the discretionary type which the operation takes as argument, and \texttt{type`} is the discretionary return type of the operation.

\subsection{Association Definitions}
\label{sec:assocdef}

\begin{nobottomsyntaxbox}
\begin{markdown}
```
association definition = class, [ qualification ], 
                         ‘-->’, [ multiplicity ], class', 
                                ‘:’, [ visibility ], variable ;
``` 
\end{markdown}
\end{nobottomsyntaxbox}

\noindent
Where \texttt{class} is the identifier of the associating object, \texttt{class'} is the identifier of the associated object and \texttt{variable} is the identifier of the instance variable that is defined by the association.

\begin{notopnobottomsyntaxbox}
\begin{markdown}
```
qualification = general map type
              | injective map type ;

general map type   = ‘"[’, type, ‘]"’ ;
injective map type = ‘"[(’, type, ‘)]"’ ; 

multiplicity = set type 
             | seq type
             | optional type ;

    set type = set0 type
             | set1 type ;
        
        set0 type = ‘"*"’ ;
    	set1 type = ‘"1..*"’ ;
``` 
\end{markdown}
\end{notopnobottomsyntaxbox}

\begin{notopsyntaxbox}
\begin{markdown}

``` 
    seq type = seq0 type 
             | seq1 type ;
        
        set0 type = ‘"(*)"’ ;
    	set1 type = ‘"(1..*)"’ ;
    
    optional type = ‘"(0..1)"’ ;
    
``` 
\end{markdown}
\end{notopsyntaxbox}
\noindent

\subsection{Access Definitions}
Access is defined as:

\begin{nobottomsyntaxbox}
\begin{markdown}
```
access member definition = [ visibility ], [ ‘static’ ] 
                         | [ ‘static’ ], [ visibility ] ;
```
\end{markdown}
\end{nobottomsyntaxbox}

\begin{notopsyntaxbox}
\begin{markdown}
``` 
visibility = ‘+’ 
	       | ‘-’ 
	       | ‘#’  ;  
``` 
\end{markdown}
\end{notopsyntaxbox}
\noindent
For public, private and protected, respectively. The default visibility to any component is private.

\section{VDM2UML Type Abstraction}
The VDM2UML type abstraction affects how compound types are represented in UML and can prevent class diagrams from becoming cluttered and verbose.
The downside to this is that the translation is no longer bidirectional since information about types may be lost. The VDM type abstraction splits VDM compound types into two groups. The groups are the primary compound types, $C_{0}$ and the secondary compound types, $C_{1}$. \\

\noindent 
$C_{0} = set,\hspace{1pt} seq,\hspace{1pt}  map,\hspace{1pt}  optional$ \\
\noindent
$C_{1} = product,\hspace{1pt} union$. \\

\noindent
Each group has a different capacity determined by $\gamma_{0}$, $\gamma_{1} \in Z*$, for $C_{0}$, $C_{1}$ respectively. The capacity determines how many compound types any given type can compose, before it is deemed too complicated for UML and therefore in need of abstraction. 
A compound type with multiple compound types within it, will belong to the group of the outer compound type. All non-basic types in the inner type count towards the capacity. If the capacity is reached, abstraction will be done in accordance to the group the type belongs to.    


\begin{nobottomsyntaxbox}
\begin{markdown}
```
abstraction = C_0 abstraction 
	        | C_1 abstraction ;

	C_0 abstraction = ‘seq of’, type_a
			        | ‘set of’, type_a
			        | ‘[’ type_a ‘]’ 
			        | ‘map’, type_a 
                    | basic type, ‘to’, type_a 
                    | basic type ;
```
\end{markdown}
\end{nobottomsyntaxbox}

\begin{notopsyntaxbox}
\begin{markdown}
```
		type_a = c_0'
               | c_1' ;

			c_0' = set...
                 | seq...
                 | [...] ;

			c_1' = ‘*’, {‘*’}
                 | ‘|’, {‘|’} ;

	C_1 abstraction = c_1' ;
```
\end{markdown}
\end{notopsyntaxbox}

\noindent
The capacity for a map type is $2\gamma_{0}$, since the map type has a minimum of two sub-types. This is also why a map type can have a basic type as one of its sub-types and still be abstracted, if the other sub-type consists of enough compound sub-types to exceed the capacity. 

For $c_1'$, the number of symbols used is given by $n-1$ where n is the number of sub-types in the non-abstracted compound type. 


\end{document}